\documentclass{JHEP3}
\usepackage{amssymb}
\usepackage{amsmath}
\usepackage{lscape}

\def\journal#1, #2, 1#3#4#5, #6{
    {\sl #1~}{\bf #2} (1#3#4#5) #6}

\def\beq{\begin{equation}}
\def\eeq{\end{equation}}
\def\ba{\begin{eqnarray}}
\def\ea{\end{eqnarray}}
\def\nn{\nonumber \\}
\def\l{\lambda}

\def\f{\varphi}

\def\e{\eta}
\def\rn{\rho_s(x)}
\def\r{\rho}
\def\p{\partial}
\def\bfl{\begin{flalign}}
\def\efl{\end{flalign}}

\newcommand{\h}{Hamiltonian}
\newcommand{\col}{collective}
\newcommand{\pv}[1]{{-  \hspace {-4.0mm} #1}}
\newcommand{\pctg} {{\cal P}\! \cot}
\newcommand{\baa}{\begin{eqnarray}}

\title{Solitons and giants in  matrix models}
\author{Ivan Andri\'c, Larisa Jonke, Danijel Jurman \\  
Theoretical Physics Division, Rudjer Bo\v skovi\'c Institute, \\
POB 180, 10002 Zagreb, Croatia \\
E-mail:\email{iandric@irb.hr},
\email{larisa@irb.hr},
\email{djurman@irb.hr}}

\abstract{
We present a method for solving BPS 
equations obtained in the collective-field approach to matrix models.
The method enables us to find BPS solutions and quantum excitations around
 these solutions in the one-matrix model, and in general for 
the Calogero model. 
These semiclassical solutions correspond to giant gravitons described by matrix models
obtained in the framework of AdS/CFT correspondence. The two-field model, 
associated with two types of giant gravitons, is investigated.
In this duality-based matrix
model we find the finite form of the 
$n$-soliton solution.  The singular limit of this solution is examined
 and a realization of open-closed string duality is proposed.
}
\begin{document}
\section{Introduction}

In the framework of AdS/CFT correspondence, the dynamics of giant gravitons
was studied using matrix model with the harmonic-oscillator potential \cite{Cor,ber,Cald}. 
The interpretation of the matrix eigenvalues as 
fermions allows a description of gravitational excitations in the 
holographic dual 
of $N=4$ SYM in terms of droplets in the phase space occupied by 
fermions.  
In particular, it was shown that giant
gravitons expanding along $AdS_5$ and $S^5$ could be interpreted as a
single excitation high above the Fermi sea, or as a hole in the Fermi
sea, respectively \cite{ber}. 
In Ref.\cite{LLM} a correspondence between the 
general fermionic droplet  and the classical
ansatz for the AdS configuration was established.
From Ref.\cite{Cor} it follows that giant gravitons are described by
particular combinations of single-trace and multi-trace operators, 
known as Schur polynomials, pictorially represented by Young diagrams.
For example, the afore-mentioned giants on
$AdS_5$ and $S^5$ are represented by two types of Young tableaux:
the one with the one-row diagram and the other with the one-column diagram \cite{Cor,ber}. 
The matrix model with the harmonic-oscillator potential is related to the free matrix model
 via su(1,1) algebra which
contains Hamitonians of both models as generators. 
As a consequence, their eigenstates are related via coherent states
or by time reparametrization \cite{Avan}.
Therefore, a detailed discussion of BPS solutions of the matrix model without 
harmonic potential is relevant for giant graviton physics.

In this paper we analyse the model without the external potential and 
present a method for obtaining BPS solutions.
Our starting point is the background-independent model defined by the action
\baa\label{a1}
S=\frac{1}{4}\int dt Tr \dot M^2(t),\ea
where $M$ is the $N\times N$ matrix.
In Ref.\cite{ant}, the analysis of the 
states of the free, $U(N)$ gauge-invariant matrix model represented by Young diagrams revealed two single-particle branches.
The one-row diagram represents a single 
particle-like excitation above the Fermi sea with the dispersion law
\beq
\label{dl1}
\omega(k)=\frac{1}{2}(k+k_F)^2-\frac{k_F^2}{2},\eeq
and the one-column diagram represents a hole-like excitation in the Fermi sea with the 
dispersion law 
\beq\label{dl2}
\omega(k)=\frac{k_F^2}{2}-\frac{1}{2}(k-k_F)^2.\eeq 
The dispersion laws (\ref{dl1}) and (\ref{dl2}) made possible the construction
of effective collective-field Lagrangians 
for each of these branches \cite{ant}, 
which are recognized as Lagrangians of $O(N)$ \cite{AB} and $Sp(N)$ \cite{AJL} gauge-invariant matrix models. 
These matrix models are connected by duality \cite{jhep1},
similarly to the duality between $S^5$ and $AdS_5$
giant gravitons \cite{Bena}.
As noted in Ref.\cite{4}, the $O(N)$ and $Sp(N)$ gauge-invariant matrix models 
have their own string interpretation as unoriented (super)strings in two dimensions.

On a singlet subspace the free matrix model (\ref{a1}) reduces to the 
quantum mechanics of the  $N$ eigenvalues $x_i$ of the matrix $M$.
Introduction of the invariant measure over the matrix configuration space into the
wavefunctions produces a prefactor $\prod_{i<j}^{N}(x_i-x_j)^{\l}$ which specifies generalized statistics
\cite{Polychronakos:1999sx},
controlled by the parametar $\l$. 
For matrix models the parameter $\l$ is related to the number of independent matrix  degrees of freedom
and takes the value $\l=1/2,1,2,$ in the case of $O(N)$, $U(N)$ or $Sp(N)$ gauge-invariant matrices, respectively \cite{jhep1}.
Finally, the dynamics of the eigenvalues is determined by the Hamiltonian 
of the Calogero type \cite{cal}
\baa\label{mqm}
H_{\rm CM}(\{x\};\l)
=-\frac{1}{2}\sum_i\frac{d^2}{dx_i^2}+\l(\l-1)\sum_{i<j}\frac{1}{(x_i-x_j)^2}.
\ea
Keeping in mind the relevance of the Calogero-type models for different branches of physics,
for example in the quantum Hall effect \cite{iso}, spin models \cite{Aniceto}, 
hydrodinamical models \cite{wieg}, systems of particles with generalized 
statistics \cite{Polychronakos:1999sx}, 
two-dimensional QCD \cite{Minahan} and black-hole physics \cite{gibb},
we use a general $\l$ throughout the paper.

In order to study the large-$N$, continuum limit of the model defined by action (\ref{a1}), 
we use the collective-field approach, developed in Ref.\cite{js}.
Introducing the collective fields $\r$ and $\pi$
\baa
&&\r(x,t)=\frac{1}{2}\int\frac{dk}{2\pi}e^{ikt}\r_k(t),\;\r_k(t)=\frac{1}{2} Tr e^{-ikM(t)},\nn
&&\left[\partial_x \pi(x), \r(y) \right]=-i\partial_x \delta(x-y),
\ea
the matrix model (\ref{a1}) was recast  in the following \h:
\baa\label{h3}
H_{\rm CM}([\r];\l)&=&\frac{1}{2}\int dx \r(x)(\p_x\pi(x))^2+
\frac{1}{8}\int dx\r(x)
\left[(\l-1)\frac{\p_x\r(x)}{\r(x)}-2\pi\l \r^H(x)\right]^2-\nn
&-&\frac{\l}{2}\int dx \left.\r(x)\p_x\frac{P}{x-y}\right|_{y=x}-
\frac{\l-1}{4}\int dx \left.\p_x^2\delta(x-y)\right|_{y=x},\ea
where $\rho^H$  is the Hilbert transform of $\r$ defined by 
$$\rho^H(x)=-\frac{1}{\pi}\pv{\int} dy \frac{\rho(y)}{x-y}.$$
The collective field $\r(x)$ can be
viewed as a bosonized 'fermionic' wave function of the afore-mentioned
discrete case. 
Another approach to bosonization, suited for a finite number of fermions 
in one space dimension was proposed in Ref.\cite{dhar1}. 

In attempt to go beyond the case of free fermions,
the authors of Ref.\cite{Don} 
start with two matrices and treat one of the matrices in the 
collective-field
theory approach, while the other is treated in the coherent-state representation.
This leads to the eigenvalue equations first found in \cite{Mar}, describing angular degrees of freedom of the 
single-matrix model.  
Another generalization of the free fermion picture was proposed 
in Ref.\cite{AAAP}, leading to the connection  between
multi-charge BPS operators of $N=4$ SYM and the
supersymmetric generalizations of the spin Calogero
system.
These developments motivated us to  
look also into the Lagrangian,
introduced in Refs.\cite{f,jhep1}, with two collective fields:
one field associated with real symmetric matrix (with $O(N)$ invariance) and
the other field associated with quaternionic matrix 
(with $Sp(N)$ invariance). This two-field model arises from the decomposition of the 
hermitian matrix into the sum of symmetric and antisymmetric matrix and 
can be thought of as
a model of interacting giant gravitons.

The plan of the paper is the following. In section 2 
we present a method for solving the BPS equation for the one-matrix model.
The construction of the conformal field enables us to reduce the problem to the
Riccati differential equation, which we relate to the Benjamin-Ono equation previously found in
\cite{ant,wieg}. Using the Riccati equation we investigate quantum excitations around the BPS solutions
and related dispersion laws.
 In section 3 we analyse the two-field model and present the form of the 
BPS solution for $n$ solitons.
 The singular limit of some of these solutions is examined and a 
 realization of open/closed string duality is proposed.
In the last section we summarize the main results and discuss some open 
questions.   

\section{One-matrix model}
\subsection{Riccati equation and boundary fields}
Here we present a method for constructing  BPS solutions of the 
\h \ (\ref{h3}).
The terms in the second line in (\ref{h3}) are singular  counter terms, 
which do not contribute in the leading order in N.
Assuming that the field $\p_x\pi(x)$ vanishes, 
the leading part of the \col-field \h\ (\ref{h3}) in the $1/N$ expansion 
is given by the effective potential 
\bfl\label{v01}
&V_{eff}=\frac{1}{2}
\int dx \rho(x)\left[\frac{\l-1}{2}\frac{\partial_x\rho(x)}{\rho(x)}
-\l\pi \r^H(x)\right]^2.
\end{flalign}
The effective potential  can be rewritten as
\bfl \label{v1}
&V_{eff}=E_0+\frac{1}{2}
\int dx \rho(x)\left[\frac{\l-1}{2}\frac{\partial_x\rho(x)}{\rho(x)}+\frac{q(1-\l)}{2}
\pctg \left(\frac{qx}{2}+\f\right)-\l\pi \r^H(x)\right]^2,
\end{flalign}
where the additional term in $V_{eff}$ is defined as
\ba \label{principal}
\pctg (qx/2+\f)=\lim_{\epsilon \to 0} \frac{\sin(qx+2\f)}{\cosh \epsilon-\cos(qx+2\f)}.
\ea
In (\ref{v1}), $E_0$ represents the terms which are to be subtracted because of the addition of
$\pctg (qx/2+\f)$ term into the square brackets:
\ba\label{e00}
E_0=&&\frac{q(\l-1)^2}{4}\int dx\p_x\rho(x)\pctg \left(\frac{qx}{2}+\f\right)-
\frac{q^2(\l-1)^2}{8}\int dx \rho(x)\pctg^2 \left(\frac{qx}{2}+\f\right)-\nn
-&&\frac{q\pi\l(\l-1)}{2}\int dx \rho(x)\rho^H(x)\pctg \left(\frac{qx}{2}
+\f\right).\ea
Assuming the
compact support $[-L/2,L/2]$, using the normalization condition $\int dx \rho(x)=N$ and the identity
\beq
(f^H g+f g^H)^H=f^H g^H-fg+f_0 g_0,\;{f_0\choose g_0}=\frac{1}{L}\int dx
{f(x)\choose g(x)},
\eeq
and performing partial integration one obtains
\ba\label{e00r}
E_0=&&\frac{qN(1-\l)}{8}\left[(1-\l)q+2\pi\l\frac{N}{L}\right]+\left.
\frac{q(\l-1)^2}{4}\rho(x)\pctg \left(\frac{qx}{2}+\f\right)
\right|_{-L/2}^{L/2}+\nn
+&&\frac{q\pi\l(\l-1)L}{4}\left.\left(\rho^2(x)-{\rho^H}^2(x)\right) \right|_{x=-2\f/q}.
\ea
At this point, $q$ and $\f$ are free parameters
to be determined by boundary conditions such that the last
two terms in (\ref{e00r}) should vanish and by 
the condition that $E_0$ should be a non-negative 
constant. 
The precise choice  of $q$ and $\f$ satisfying these conditions 
determines different solutions of the model 
and is discussed in the next subsection.
Here we proceed with the analysis of the effective potential (\ref{v1}). 
With $E_0$ being a constant, the  contribution of $V_{eff}$ to the
\h \ is minimized by a solution of the
integro-differential Bogomol'nyi-type equation
\beq \label{b1}
\partial_x\rho=q \pctg \left(\frac{qx}{2}+\f\right)\r+\frac{\l\pi}{\l-1}2\r \rho^H.
\eeq 
Taking the Hilbert transform of Eq.(\ref{b1}),
we find the equation for $\r^H$,
\bfl\label{r2}
\partial_x \rho^H=q \pctg \left(\frac{qx}{2}+\f\right)\rho^{H}-q\r_0
-\frac{\l\pi}{\l-1}\left(\rho^2-{\rho^{H}}^2-\rho_0^2 \right).
\end{flalign}
We construct the conformal field $\Phi$ containing only the positive frequency part of $\rho$ 
\beq
\Phi=\rho^H+i\rho=\frac{1}{\pi}\int dz \frac{\r(z)}{z-x-i\epsilon},
\eeq
and satisfying the Riccati differential equation
 \beq\label{rc1}
 \partial_x\Phi=\frac{\l\pi}{\l-1}\Phi^2+q \pctg \left(\frac{q x}{2}+\f\right)\Phi
+\frac{\l\pi\rho_0^2}{\l-1}-q\r_0.\eeq
The physical interpretation of the Riccati differential equation can be found from the relation to
the Benjamin-Ono equation \cite{wieg}. Taking into account the 
Bogomol'nyi limit, we 
evaluate the fields $u^+$ and $u^-$ from Ref.\cite{wieg}:
\beq\label{uovi}
u^-=\sqrt{\l}\pi \Phi,\;u^+=\frac{\l-1}{\sqrt{\l}} \frac{q}{2} \pctg \left(\frac{q x}{2}+\f\right)
\eeq
and then plugging (\ref{uovi}) into the 
Benjamin-Ono equation, we obtain Eq.(\ref{rc1}).

Obviously, if the field $\r$ is a solution of Eq.(\ref{b1}), 
then the field $\Phi$ necessarily
satisfies Eq.(\ref{rc1}). The converse is not true in general, and to obtain 
the field $\r$
from the solution of the Riccati equation (\ref{rc1}), 
it is sufficient that the condition
\beq\label{condition}
\Phi^H(x)=i\Phi(x)+\rho_0
\eeq
holds. 
In this case, the  solution of Eq.(\ref{b1}) is simply given 
as $\rho=-i(\Phi-\Phi^*)/2$.
Equation (\ref{rc1}) can be further transformed into the second-order 
Schr\"odinger-like differential equation by making the substitution 
$\Phi=(1-\lambda) \partial_x v/(\l \pi v)$:
\ba \label{ujed}
\partial_x^2 v=q \pctg \left(\frac{qx}{2}+\f \right) \partial_x v-\frac{\l\pi\r_0}{1-\l}
\left(\frac{\l\pi\r_0}{1-\l}+q\right)v.
\ea
\subsection{Semiclassical solutions}
The construction of the Riccati equation enables us to obtain the semiclassical static solutions.
Generally, there are two methods for solving the Riccati equation. 
One method is to construct a general solution from the 
known particular solution, while the other method 
is based on solving the second-order linear differential equation (\ref{ujed}).
Applying both methods, 
we find some interesting solutions, given in Table \ref{tabla1}.

In the following analysis we differentiate two possibilities: $\l<1$ and $\l>1$.
First, we discuss the case $\l<1$. 
The particular solution of the Riccati equation (\ref{rc1}) for non-vanishing $\rho_0$
is given in the first row of Table \ref{tabla1}.
The parameter of this solution satisfies the relation 
 \beq\label{n2}
e^t=1+\frac{q(1-\l)}{\l\pi \r_0}.\eeq
Taking into account the boundary conditions
\beq\label{n1}
\rho^H(-\frac{2\f}{q})=\rho(-\frac{2\f}{q})=0,\;
\pctg (\frac{qL}{4}+\f)=0,\eeq 
we find 
\beq\label{slijediku}
q=2\pi M/L,\;M \in \mathbb{N},
\eeq
where the number $M$ can be interpreted as the number of solitons.
In order to have odd $M$, we take $\f=0$, whereas for even $M$ we take $\f=\pi/2$.
Taking into account the normalization condition, we find 
\beq\label{n3}
e^t=1+\frac{2M(1-\l)}{N\l}.\eeq
From the $M$-soliton solution in the limit $L\rightarrow \infty$, 
keeping $\r_0$ fixed and defining 
\beq\label{n4}
b=(1-\l)/(\l\pi \r_0),\eeq
 we find
the one-soliton solution ($M=1$, $\f=0$) obtained in Refs.\cite{poly,plb}.
The energy is just the energy of one soliton obtained by taking the 
corresponding limit. 
The uniform zero-energy solution $\r(x)=\r_0$ is obtained in the limit $q \to 0$, taking $\f=\pi/2$.\\
\begin{table*}
\begin{tabular}{cccc}
$\l$ & $\Phi_s(x)$ & $\r_{s}(x)$ & $E_0$\\
\hline
$\l<1$ & $\frac{iq(1-\l)}{\l\pi(e^{t}-1)}\frac{1-e^{i(qx+2\f)}}{1- e^{-t} e^{i(qx+2\f)}}$ &
$\frac{q(1-\l)\coth (t/2)}{2\pi\l}\frac{1-\cos(qx+2\f)}{\cosh t-\cos(qx+2\f)}$ & 
$\frac{(1-\l)\pi^2}{2L^2}\scriptstyle{[\l N^2 M+(1-\l) N M^2]}$\\
 & $\frac{1-\l}{\l\pi b}\frac{i x}{x+ib}$ & $\frac{1-\l}{\l\pi b}\frac{x^2}{x^2+b^2}$ & 
$\frac{(1-\l)^3}{2\l b^2}$\\
\hline
$\l>1$ & $\frac{ik(\l-1)}{2\pi\l}\frac{1+e^{-t}e^{ikx}}{1-e^{-t}e^{ikx}}$ & $\frac{k(\l-1)}{2\pi\l}\frac{\sinh t}{\cosh t-\cos k x}$ & $0$\\
 & $\frac{1-\l}{\l\pi}\frac{1}{x+ib}$ & 
$\frac{\l-1}{\l\pi}\frac{b}{x^2+b^2}$ & $0$\\
\hline
$\l \lessgtr 1$ & $i\r_0$ & $\r_0$ & 0\\
\end{tabular}
\caption{\label{tabla1}BPS solutions}
\end{table*}
\indent Next we discuss the  case $\l>1$. 
We take $q=0$, $\f=\pi/2$, thus eliminating  the term  $\pctg$ from 
Eq.(\ref{rc1}), and obtain the general solution of 
the Riccati equation (\ref{rc1}) for 
\beq\label{n5}
\r_0=\frac{(\l-1)k}{2\pi\l},\;k=2\pi M/L.\eeq
It is given in the third row of Table \ref{tabla1}, where $t$  
is a non-negative free parameter.
An additional solution could be obtained from the general solution 
in the limit $L\rightarrow \infty$,
 taking $t=2\pi b/L$.
In the case $t\to \infty$, we obtain the constant density solution $\r(x)=\r_0$. 
Taking into account the normalization condition  we obtain that the 
number of solitons
$M$ exceeds the number of particles $N$ giving us the  relation
 \beq\label{nennorm}
\l=M/(M-N).
\eeq
\indent Solitons on the compact support from Table \ref{tabla1} are of the same 
shape as solitons in the Sutherland model \cite{BAL}, thus reflecting the fact that the 
two models are 
interrelated via the periodicity condition. 
Using the condensed-matter language, we interpret 
the M-soliton solutions as soliton trains. In the large-$M$ limit, these solutions can be viewed as crystal-like structures
with periodicity $2\pi/q$ or $2\pi/k$. The one-soliton solutions are regarded 
as composite particles (quasi-particles).
Resembling the holes (lumps), these one-soliton solutions enable us to
identify the M-soliton solutions  found for $\l<1$ ($\l>1$) as 
dark (bright) solitons.
In Ref.\cite{plb} it was shown that adding the term $(1-\l)/(x-z)$ into the effective potential
was equivalent to the extraction of the
prefactor $\prod_i(x_i-z)^{1-\l}$ from the wave
function of the Hamiltonian (\ref{mqm}).
This equivalence enables us to associate a quasi-particle 
located at $z$ with the prefactor of the wave
function.
Consequently, the additional term $\pctg (qx/2+\f)$ is associated with 
the prefactor describing $M$ equidistant quasi-particles.

The soliton solutions we have found  in the
collective-field formulation of the free matrix model correspond to the particle and hole states 
in the system  (\ref{mqm}) of nonrelativistic fermions \cite{ant}.
Owing to the $su(1,1)$ dynamical symmetry \cite{jhep1}, the eigenstates of the \h \  (\ref{mqm})
can be represented as generalized coherent states 
of the same \h \ with the additional harmonic potential interaction between fermions \cite{AFF,P,GMP}.
The particle and hole states 
in the system of fermions with the harmonic potential interaction
correspond to the giant gravitons of a $1/2$ BPS sector of $N=4$ SYM \cite{Cor,ber,Cald,LLM}. 
Therefore, our solutions correspond to the coherent states of the 
matrix model with the harmonic potential, i.e. to the quasi-classical CFT duals 
of the giant gravitons in AdS constructed in Ref.\cite{Cald}.
The nonexistence of the quasi-classical CFT dual of the single giant graviton on the sphere $S^5$ is reflected throught the
relation (\ref{nennorm}), from which it follows that the soliton with $M=1$ in the $\l>1$ case is non-normalizable
since $M$ must exceed $N$, in accordance with the conclusion of Ref.\cite{Cald}.

At this point we would like to emphasize a simple relation between systems 
with $\l<1$ and those with $\l>1$.
By substituting $\l \rho(x)=\alpha-m(x)$ into Eq.(\ref{b1}) for $\l>1$ 
(without the term $\pctg$)
and by inserting explicit forms of the solutions for the term  $\r^H/\r$,
  we find
that the field $m$ satisfies Eq.(\ref{b1}) 
for $\l'=1/\l<1$ (with the  term $\pctg$). 
This agrees with the result obtained in Ref.\cite{PolychronakosMinahan} 
in the k-space ($\rho_k \rightarrow -m_k/\l$). 
The  difference is that our relation is valid for the BPS equations,
whereas in  Ref.\cite{PolychronakosMinahan} the 
duality relation connects quantum \h s.
Recall that the solution of the form similar to that given in the third row of 
Table \ref{tabla1} was found as a solution of the dynamic
 equations of motion of the 
Calogero model in Ref.\cite{poly}.
This signalizes that there is a generalization of our method for dynamic 
equations of motion (some progress has been made in Ref.\cite{wieg}).
Finally, we stress that the construction of the Riccati equation 
is possible for the model trapped in the harmonic well, providing a
new method for analysing this model.

\subsection{Quantum excitations around semiclassical solutions}
To get an insight into the dynamics of quantum excitations,
we expand the Hamiltonian (\ref{h3}) around the semiclassical solution
\beq\label{eta1}
\rho (x,t)=\rho_s(x)+\partial_x\eta(x,t),\eeq 
where $\e$  is a small density quantum 
fluctuation around the soliton solution $\rho_s$ of Eq.(\ref{b1}).
The quadratic part of the Hamiltonian can be written in the following form:
\beq\label{hw}
H^{(2)}=\frac{1}{2}\int dx\rho_s(x)A^{\dagger}(x)A(x),
\eeq
where we have introduced the operators $A$ 
\ba\label{op}
A&=&-\pi_\eta+i\left[\frac{(\l-1)}{2}\partial_x\frac{\partial_x\eta}{\r_s}
-\pi\l\partial_x \eta^H\right],
\ea
satisfying the following equal-time commutation relation:
\ba
\left[A(x),A^{\dagger}(y)\right]=(1-\l)\p^2_{xy}\frac{\delta(x-y)}{\rn}+2\l
\p_x\frac{P}{x-y}.\label{comma} \ea
Using the equation of motion $\dot A(x,t)=i[H,A(x,t)]$, we obtain the equation 
\ba\label{eq2}
\left[-i\p_t+\frac{\l-1}{2}\frac{\p_x\r_s}{\r_s}\p_x-\frac{\l-1}{2}\p_x^2\right](\r_s A)
=-\l\pi\r_s\p_x (\r_s A)^{H}.
\ea
Taking the Hilbert transform of this equation, and using Eq.(\ref{b1}), we find
\ba\label{eq2h}
\left[-i\p_t +\frac{\l-1}{2}\frac{\p_x\r_s}{\r_s}\p_x-\frac{\l-1}{2}\p_x^2\right](\r_s A)^H
=\l\pi\r_s\p_x (\r_s A).
\ea
Defining the fields
\beq\label{ff}
\Phi_s^{\pm}=\r^H_s\pm i\r_s,\;
\phi^{\pm}=(\r_s A)^H \pm i(\r_s A),
\eeq
we find that $\phi^\pm$ satisfies 
\bfl \label{eqf}
\left\{ i\p_t -
 \left[\l \pi \Phi_s^{\pm}+\frac{q(\l-1)}{2}\pctg \left(\frac{qx}{2}+\f\right)\right]\partial_x+
\frac{\l-1}{2}\p_x^2\right\}\phi^\pm=0.
\end{flalign}
This equation can be obtained from the 
Riccati equation (\ref{rc1}) by adding 
$2i\p_t \phi^{\pm}/(\l-1)$ on the LHS and expanding $\Phi(x,t)$ around 
the solution 
$\Phi_s(x)$, $\Phi^{\pm}(x,t)=\Phi_s^{\pm}(x)+\partial_x \phi^{\pm}(x,t)$, 
keeping only  terms linear in $\phi$. 
Therefore, one can interpret the field $\phi$ as a fluctuation around
the conformal field $\Phi_s$.
Solving Eq.(\ref{eqf}) for the solutions found in section 3, we obtain the 
following results:\\
-the operator A is given by
\ba\label{razvojA}
A=\frac{2\pi}{L}\sum_{n,s} e^{i\omega_n t} f_{n,s}(x) \left[\theta (\omega_n) a_{n,s}+\theta (-\omega_n) a_{n,s}^{\dagger}\right],
\ea
where the operators $a_{n,s}$ satisfy 
\beq
[a_{n,s},a^{\dagger}_{m,s'}]=|\omega_n| L/\pi \delta_{nm}\delta_{ss'}
\eeq
 and 
the functions $f_{n,s}$ are orthonormalized with respect to the 
measure $\r_s(x)$ as follows:
\ba\label{Norma}
\int_{-L/2}^{L/2} dx \r_s(x) f_{n,s}^*(x) f_{m,s'} (x)=\frac{L}{2\pi}\delta_{nm}\delta_{s,s'};\ea
-the Hamiltonian up to quadratic terms is given by
\ba\label{rezovi}
H=E_0+\frac{\pi}{L} \sum_{n,s} a_{n,s}^{\dagger} a_{n,s}+\sum_{n,s} \theta (-\omega_n) |\omega_n|.
\ea  
The functions  $f_{n,s}$ and the eigenvalues $\omega_n$ for all solutions found in section 3 are given in 
Table \ref{tabla2}, where $k_0=| \l \pi \r_0/ (\l-1) | $.

Comparing the known dispersion laws (\ref{dl1}), (\ref{dl2}) for 
the quantum excitations on the uniform background, 
obtained from the Young diagrams in Ref.\cite{ant} with 
the dispersion laws given in Table \ref{tabla2} we find an agreement.
At the semiclassical level we obtain that the system can be in the 
fluid phase (uniform density)
as well as in the crystal-like configuration.
So, it would be interesting to calculate the phase transition amplitudes
and also to calculate the correlation functions determined 
in this approach by the quadratic Hamiltonian.
\newpage
\begin{landscape}
\begin{table*}
\begin{tabular}{ccrll}
$\l$ & $\r_s$ & $f_{n,\pm}$ & & $\omega_n$\\
 \hline
 & $\frac{q(1-\l)\coth (t/2)}{2\pi\l}\frac{1-\cos(qx+2\f)}{\cosh t-\cos(qx+2\f)}$ & 
$\scriptstyle{\sqrt{\frac{\l (k_0+q)(k_n+q)}{4(1-\l)k_0 k_n (2k_0+q)}}}\left(\scriptstyle{1}-\frac{k_ne^{\pm i(qx+2\f)}}{k_n+q}\right)\!\!
\left(\scriptstyle{1}-\frac{k_0 e^{\mp i(qx+2\f)}}{k_0+q}\right)\frac{e^{\pm i(k_n-k_0)x}}{1-\cos (qx+2\f)}$ & 
$\scriptstyle{k_n>k_0}$ &
$\!\frac{1-\l}{2}\scriptstyle{(\!k_n\!+k_0+q\!)(\!k_n-k_0\!)}$\\
$\scriptstyle{\l<1}$ & $\frac{1-\l}{\l\pi b} \frac{x^2}{x^2+b^2}$ &
$\scriptstyle{\sqrt{\frac{\l}{2k_0(1-\l)}}\left(1\pm \frac{i}{k_n x}\right)\left(1\mp \frac{i}{k_0 x}\right)\;e^{\pm i(k_n-k_0)x}}$ & $\scriptstyle{k_n>k_0}$ & $\frac{1-\l}{2}\scriptstyle{\left(k_n^2-k_0^2\right)}$\\
 \hspace{-1cm} & $\scriptstyle{\r_0}$ & $\scriptstyle{\frac{1}{\sqrt{2\pi\r_0}}e^{\pm i(k_n-k_0)x}}$ & $\scriptstyle{k_n>k_0}$ & $\frac{1-\l}{2}\scriptstyle{\left(k_n^2-k_0^2\right)}$\\
\hline
 & $\frac{k(\l-1)}{2\pi\l}\frac{\sinh t}{\cosh t-\cos k x}$ & 
$\scriptstyle{\sqrt{\frac{\l}{2k_0(\l-1)\left(1-e^{-2t}\right)}}\left(1- e^{-t}\;e^{\mp 2ik_0x}\right)\;e^{\pm i(k_n+k_0)x}}$ & $\scriptstyle{k_n>-k_0}$ & $\frac{\l-1}{2}\scriptstyle{\left(k_0^2-k_n^2\right)}$\\
 $\scriptstyle{\l>1}$ & $\frac{\l-1}{\l\pi}\frac{b}{x^2+b^2}$ &
$\scriptstyle{\sqrt{\frac{\l}{2b(\l-1)}}(x\mp ib)}\;e^{\pm ik_n x}$ & $\scriptstyle{k_n>0}$ & $-\frac{\l-1}{2}\scriptstyle{k_n^2}$\\
 & $\scriptstyle{\r_0}$ & $\scriptstyle{\frac{1}{\sqrt{2\pi\r_0}}e^{\pm i(k_n+k_0)x}}$ & $\scriptstyle{k_n>-k_0}$ & $\frac{\l-1}{2}\scriptstyle{\left(k_0^2-k_n^2\right)}$\\
\end{tabular}
\caption{\label{tabla2}Excitations around BPS solutions}
\end{table*}
\end{landscape}

\newpage
\section{Duality-based matrix model}
\subsection{Semiclassical solutions}

In attempt to go beyond the case of free fermions, in this section we analyse a 
generalization of the hermitian matrix model, introduced in Refs.\cite{f,jhep1}.
Actually, this model was first formulated in Ref.\cite{pra} as a duality-based generalization of the Calogero model, 
and is defined by the Hamiltonian
\bfl\label{z}
\!H(x,z)\!=&\!\!\!\sum_{i=1}^N \frac{p_i^2}{2}+\!\frac{1}{2}\!\sum_{i\neq j}^N\!\!
\frac{\l(\l-1)}{(x_i-x_j)^2}+\!\frac{1}{2}\!
\sum_{i,\alpha}^{N,M}\!\!\frac{(\kappa\!+\!\l)(\kappa\!-\!1)}{(x_i-Z_{\alpha})^2}+\nn
\!+&\frac{\l}{\kappa}\left[\sum_{\alpha=1}^M \frac{p_{\alpha}^2}{2}+\frac{1}{2}
\sum_{\alpha\neq\beta}^M
\frac{\kappa^2/\l\left(\kappa^2/\l-1\right)}{(Z_{\alpha}-Z_{\beta})^2}\right]
.\end{flalign}
In Ref.\cite{jhep1}, it was shown that 
for $\l=1/2$ this model arises 
from the decomposition of the
hermitian matrix into the sum of symmetric and antisymmetric matrix.
Transformation into the hydrodynamic formulation, in terms of the density operators $\rho(x)$, $m(x)$ and 
the corresponding conjugate operators
for $\kappa=1$, results in the hermitian \col-field \h ~\cite{jhep1}
\bfl\label{zc}
H&=\frac{1}{2}\int dx \r(x)(\p_x\pi_{\r}(x))^2+\frac{\l}{2}\int dx m(x)(\p_x\pi_{m}(x))^2+\nn
&+\!\!\int\!\! dx\frac{\r(x)}{2}
\left[\frac{\l-1}{2}\frac{\p_x\r(x)}{\r(x)}+\pv\int dy \frac{\l\r(y)}{x-y}+
\pv\int dy\frac{m(y)}{x-y}\right]^2\!\!+\nn
&+\!\!\int\!\! dx\frac{m(x)}{2\l}
\!\!\left[\frac{1-\l}{2}\frac{\p_x m(x)}{m(x)}+\pv\int dy
\frac{m(y)}{x-y}+
\pv\int dy\frac{\l\r(y)}{x-y}\right]^2\!\!\!-\nn
&-\frac{\l}{2}\!\!\int\!\! dx \r(x)\p_x\frac{P}{x-y}\underset{x=y}{|}\!\!-
\frac{1}{2}\!\!\int\!\! dx m(x)\p_x\frac{P}{x-y}\underset{x=y}{|}\!.
\end{flalign}
The terms in the last line in  (\ref{zc}) are singular  counter terms, 
which do not 
contribute in the leading order in $N$ and $M$.
This model was analysed in Ref.\cite{jhep2}, where
some solutions were constructed.
Here, we use the Riccati equation to
construct some new solutions in this  duality-based  model.
We are looking for the solutions of two coupled Bogomol'nyi equations,
\ba\label{gs1}
&&(\l-1)\partial_x\rho-
2\pi\rho(\l\rho^H+m^H)=0,\\
\label{gs2}
&&(1-\l)\partial_x m-
2\pi m(\l\rho^H+m^H)=0.
\ea
Based on the duality, we make an ansatz $m^H=-\l \alpha \rho^H/\rho$.
Equation (\ref{gs1}) now becomes
\beq\label{gsa1}
(\l-1)\partial_x\rho-2\l\pi\rho\rho^H+2\l \alpha\pi\rho^H=0.\eeq
Following the method from  section 2,  
we construct the field $\Phi=\rho^H+i\rho$ which satisfies the 
following Riccati equation:
\beq \label{riczadu}
\p_x \Phi=\frac{\l \pi}{\l-1} \Phi^2-i\frac{2\l \pi\alpha }{\l-1}\Phi+
\frac{\l \pi \rho_0}{\l-1}(\rho_0-2\alpha).
\eeq 
The general solution of this equation constructed from the constant solution $\Phi=i\rho_0$ is
\beq \label{generalfi} 
\Phi(x)\!=i\rho_0- \frac{\l-1}{\l \pi}\frac{iqce^{iqx}}{1+ce^{iqx}},\;q=\frac{2\l \pi(\alpha-\r_0)}{1-\l}>0.\eeq
The solutions for $\r$ and $m$ $(c=e^{i\phi-u-v},\;|c|<1)$ are
\bfl\label{spec2}
&\rho(x)=\alpha\frac{\cosh (u-v)+\cos (qx+\phi)}{\cosh (u+v)+\cos (qx+\phi)},\;m(x)=\frac{\tilde{c}}{\rho(x)},\\
&q=\frac{4\l \pi \alpha}{1-\l} \frac{\sinh u \sinh v}{\sinh (u+v)}
,\;\frac{\tilde{c}}{\l \alpha^2}=\frac{\sinh (u-v)}{\sinh (u+v)},\;u>v>0\nonumber.
\end{flalign}
Taking into account the normalization conditions and the 
compact support $[-L/2,L/2]$, $q=2\pi n/L$,
we find the following relations:
\ba \label{nabac}
&&\r_0=\frac{N}{L},\;  \alpha=\frac{\l N+(1-\l)n}{\l L},\;\coth u=2+\frac{\l N-M}{(1-\l)n},  
\nn
&&m_0=\frac{M}{L},\;\frac{\tilde{c}}{\alpha}=\frac{M-(1-\l)n}{L},\;\coth v=\frac{\l N +M}{(1-\l)n}.\ea
From the solution (\ref{spec2}) taking $\phi=\pi,\;\sinh (u/2-v/2)=aq/2,\;\sinh (u/2+v/2)=bq/2,$ $b>0$, 
 and taking the limit $q\rightarrow 0$,
we obtain the one-soliton solution
\bfl\label{s0c}
\rho(x)=\alpha \frac{x^2+a^2}{x^2+b^2},\; m(x)=\frac{\lambda \alpha^2 a}{b\rho(x)},\;a^2=b^2+\frac{\l-1}{\l\pi\alpha}b.
\end{flalign}
Here we would like to emphasize that the properties of the solutions of 
the dual model (\ref{zc}) resemble the properties of the dual giant gravitons 
on AdS ($\l<1$) and on the sphere ($\l>1$) from Refs.\cite{Bena,Cor,ber}.

\subsection{Semiclassical solutions in the singular limit} 

In this subsection we discuss the existence of the singular
solutions of the duality-based model and the methods for finding them.
Inspired by the one-soliton solution of the model introduced in \cite{f}, we proposed the existence of the
multi-soliton solutions in the same paper. 
In the paper \cite{jhep2} we discussed these solutions in the context of the 
new 
matrix-model interpretation of the duality-based Hamiltonian. 
We started from an assumption that there exists the 
finite form of a multi-soliton solution.
In the case of the one-soliton solution, 
the finite form of the solution was known 
and we noticed that, in the singular limit, 
the conditions on the parameters of the solution were reduced.
This indicates that in the singular limit, the Bogomol'nyi equations  
are less sensitive to the details of the precise form of the finite solution.
In the hope of getting a hint about the 
finite form of a multi-soliton solution, 
we used a simple ansatz which in itself was not a solution in its finite form,
but had the same singular limit as the finite solution.
We found that the ansatz solved the equations in the singular limit.
Although the calculation of the contribution to the Hamiltonian leads to an ambiguity in order of 
taking limits, this problem can be avoided by introduction of the finite support.
Now, we would like to confirm the 
existence of singular solutions and the correctness 
of the calculations performed in the paper \cite{jhep2} 
by using the finite form of the solution constructed in subsection 3.1.
Taking the limit $u-v=2\epsilon\to 0$ of the solution  (\ref{spec2}), we find 
\[\r(x)=\alpha \frac{\cos^2(\frac{qx+\phi}{2})}{\sinh^2 v+\cos^2(\frac{qx+\phi}{2})},\;
\alpha=\frac{(1-\l)q}{2\l\pi}\coth v,\]
\[
m(x)=(1-\l)\sum_{i=-\infty}^{\infty}\delta(x-x_i),\;x_i=\frac{(2i+1)\pi-\phi}{q}.\]
Using the product representation for the trigonometric function we find that 
the $\infty$-soliton solution 
exactly matches the form proposed in our paper \cite{jhep2}.
Furthermore, by taking the compact support of lenght L,
 we find the finite number of solitons.
In the case that $q$ is small, we find that this solution reduces to the 
rational ansatz, 
as proposed in papers \cite{f,jhep2}.      

In the comment made by V. Bardek and S. Meljanac in \cite{x} regarding  this part 
of our  published work \cite{jhep2}, the authors claim that such solutions do not exist.
The authors based their claims on the symbolic manipulation with delta functions 
assuming the independece of the regularization procedure. 
As we are aware that solving nonlinear coupled equations is sensitive to the 
regularization used, the application of the symbolic manipulation is not 
suitable and does not give a unique 
answer because the product of distributions is not uniquely defined.
The authors also claim that their paper \cite{xx} gives convicing arguments
 that a singular form of the solution
does not exist, but they simply have not checked the limit $c\rightarrow 0$ 
in that paper.

\subsection{Open-closed string duality?}
Solutions of the equations obtained in the collective-field formulation 
of matrix models are interpretated as giant gravitons. 
Although the existence of multi-soliton solutions in their finite form
is also an interesting question from the point of view of integrable systems,
the limiting form when one of the fields behaves as a sum of delta functions deserves special attention from the point of view of matrix theory.
Namely, in this singular configuration, it is reasonable to describe the 
$m(x)$ part of the 
Hamiltonian by a discrete set of variables.
Furthermore, we can interpret the continuum part of the Hamiltonian as 
closed string theory and the discrete part of the Hamiltonian as open string 
theory \cite{ber}. 
So, by studying the relation between these two Hamiltonians we might get 
a better insight into the open-closed string duality. 

The matrix model (\ref{a1}) (the Calogero model) possesses the 
symmetry of the action generated by the operators closing the $su(1,1)$ algebra
~\cite{pra,jhep1}:
$$[T_+,T_-]=-2T_0,\;[T_0,T_\pm]=\pm T_\pm,$$
The generators of the algebra in the discrete case are 
\baa\label{tdisc}
T_+(\{x\};\l) &=& \frac{1}{2}\sum_{i=1}^N\p_i^2+\frac{\l}{2}\sum_{i\neq j}^N
\frac{1}{x_i-x_j}(\p_i-\p_j),\nn
T_0 (\{x\};\l)&=& -\frac{1}{2}\left(\sum_{i=1}^N x_i\p_i+E_0\right),
\;T_-(\{x\};\l)=\frac{1}{2}\sum_{i=1}^N x_i^2 ,\ea
and in the continuous, large-N case,
\baa\label{tcon}
T_+([\r];\l) &=& -\frac{1}{2}\!\int \!dx \r(x)(\p_x\pi(x))^2\!+\!\frac{1}{2}\!\int \!dx
\left((\l-1)\p_x\r(x)+2\l\r(x)\pv\int dy\frac{\r(y)}{x-y}\right)
\!\p_x\pi(x),\nn
T_0 ([\r];\l)&=& -\frac{1}{2}\left(i\int dx x\r(x)\p_x\pi(x)+E_0
\right), \;T_-([\r];\l)=\frac{1}{2}\int dx x^2 \r(x),
\ea 
where the ground-state energy is $E_0(N,\l)=(\l N(N-1)+N)/2$.

Furthermore, it was shown in ~\cite{pra,f,ajur,jhep1} that there existed
 a strong-weak coupling duality in the Calogero model
and here we briefly review the main results.
In the following we use the abbreviation $[\bullet]$ for the arguments of 
operators, depending on
 which case is under consideration (discrete or continuous) and analogously for
the dual system $[\circ]$.
We also need the following definitions:
 \begin{equation}\label{jaci}
\ln J([\bullet],\l)\!=\!\!\left\{
\begin{array}{cc}
\l \sum_{i\neq j}\ln(x_i-x_j) & {\rm discrete}\\
(\l-1)\int dx \r(x)\ln \r(x)+\l
\int\int dx dy \r(x)\ln|x-y| \r(y) & {\rm continuous},
\end{array}
\right. 
\end{equation}
\baa\label{newk}
\ln V([\bullet],[\circ])\!=\!\!\left\{
\begin{array}{cc}
\sum_{i, \alpha}\ln(x_i-z_\alpha) & {\rm discrete-discrete}\\
\lim_{\varepsilon\to 0}\int\int dx dz \rho(x)
\ln(x-z-i\varepsilon)m(z) & {\rm continuous-continuous}\\
\lim_{\varepsilon\to 0}\sum_{\alpha}
\int dx \rho(x)
\ln(x-z_\alpha-i\varepsilon) & {\rm continuous-discrete}.
\end{array}
\right. 
\ea
The strong-weak duality is displayed as follows:
\baa\label{dul}
T_0([\bullet],\l)V([\bullet],[\circ])&=&\left\{-T_0([\circ],1/\l)
-\frac{1}{2}[NM+
E_0(N,\l)+E_0(M,1/\l)]\right\}V([\bullet],[\circ]),\nn
T_+([\bullet],\l)V([\bullet],[\circ])&=&-\l T_+([\circ],{1}/{\l})
V([\bullet],[\circ]).\ea
The operators $T_+([\bullet],\l)$ are equivalent to the Hamiltonians (\ref{mqm}), (\ref{h3}) 
\beq
H_{CM}([\bullet],\l)=-J^{\frac{1}{2}}([\bullet],\l)T_+([\bullet];\l)J^{-\frac{1}{2}}([\bullet],\l)
\eeq
while $T_0([\bullet],\l)$ are equivalent to the same 
Hamiltonians with an additional harmonic-well potential 
(Calogero-Sutherland Hamiltonians):
\baa
H_{CS}([\bullet];\l,\omega)&=&-2\omega J^{\frac{1}{2}}([\bullet],\l) S([\bullet];\l,\omega) 
T_0([\bullet];\l) S^{-1}([\bullet];\l,\omega)J^{-\frac{1}{2}}([\bullet],\l)=\nn
&=& H_{CM}([\bullet],\l)+\omega^2 T_-([\bullet];\l),
\ea
where
\beq
S([\bullet];\l,\omega)=e^{-\omega T_-([\bullet];\l)} e^{-\frac{1}{2\omega} T_+([\bullet];\l)}.
\eeq
Conseqently, we interpret 
\baa\label{hd2}
H_{CM}([\circ];{1}/{\l})\equiv -J^{\frac{1}{2}}([\circ];{1}/{\l})T_+([\circ];{1}/{\l})
J^{-\frac{1}{2}}([\circ];{1}/{\l})
\ea
and
\baa
H_{\!CS}\!\left([\circ];\!{1}/{\l},\!{\omega}/{\l}\right)\!\!&\equiv&\!\!-\!\frac{2\omega}{\l} 
J^{\!\frac{1}{2}}\!\!\left([\circ],\!{1}/{\l}\right)\! S\!\left([\circ];\!{1}/{\l},\!{\omega}/{\l}\right)\! 
T_0\!\!\left([\circ];\!{1}/{\l}\right)\! S^{-\!1}\!\!\left([\circ];\!{1}/{\l},\!{\omega}/{\l}\right)
\!J^{-\!\frac{1}{2}}\!\!\left([\circ],\!{1}/{\l}\right)\!\!=\nn
&=& H_{CM}\left([\circ],{1}/{\l}\right)+\frac{\omega^2}{\l^2} T_-\left([\circ];{1}/{\l}\right),
\ea
as  \h s of the dual system.
In the papers \cite{pra,f,ajur} the duality relations were used for 
construction of the spectrum generating algebra
for the two-family model, while here we show that they
 can be used for construction of the eigenstate of one system from 
the known eigenstate of the dual system.
Suppose that we know  the eigenstate $\psi_{\{E\}}$  with energy $E$, satisfying
\baa\label{eg1}
H_{CS}([\bullet];\l,\omega)\psi_{\{E\}}=E\psi_{\{E\}}.\ea
Then, the eigenstate $\phi_{\{\tilde{E} \}}$ of the dual system, with energy $\tilde{E}$,
\baa\label{eg2}
H_{CS}\left([\circ];{1}/{\l},{\omega}/{\l}\right)\phi_{\{\tilde{E} \}}=\tilde{E}\phi_{\{\tilde{E}\}},\ea
is given by
\baa\label{vv1}
\phi_{\{\tilde{E} \}}=J^{\frac{1}{2}}\left([\circ],{1}/{\l}\right)
e^{-\frac{\omega}{\l}T_-\left([\circ];{1}/{\l}\right)}
\int _{\cal{V}} \psi_{\{E\}} J^{\frac{1}{2}}([\bullet],\l) e^{-\omega T_-([\bullet];\l)}V([\bullet],[\circ]),
\ea
where the integration is performed over the corresponding configuration space and 
\beq
\l\tilde{E}+E=\omega[NM+E_0(N,\l)+E_0(M,1/\l)].
\eeq
The corresponding dual relation also holds for the Hamiltonians $H_{CM}$ because
 the 
eigenstates of these 
are realized as coherent states of the $H_{CS}$ \cite{AFF,P,GMP}, or in another approach by the use of unconventional separation of variables in the Schr\"odinger problem \cite{Jac}.
Now, the quantum mechanics (\ref{mqm}) of the eigenvalues $x_i$
of $M$ was regarded as an open string/D-brane description of the
corresponding string theory \cite{str1}. 
On the other hand, the collective field
defined in the large-$N$ limit of the
matrix model (\ref{hd2}) 
represents the closed string excitations~\cite{aj}.
Inspired by this interpretation we
propose an explicit realization of open-closed string duality.
The relation (\ref{vv1}) tells us how to construct a wave functional of
closed string excitations described
by the collective field $\r$ from the wave function of dual $M$ open string excitations,  and vice versa.

\section{Conclusion}

Motivated by the  relevance of soliton solutions in matrix models
for giant graviton physics, we have addressed the problem of existence 
of these  solutions  and have analysed their properties.
We have introduced a powerful method for obtaining these BPS solutions in the 
collective-field approach. 
The method is based on the construction of  a boundary conformal field 
out of the density of eigenvalues of a matrix, satisfying the Riccati differential equation.
This method extends to the related Calogero models, in particular to the model with 
the harmonic-well potential.
Furthermore, we have established the relation between the hydrodynamic Benjamin-Ono equation 
and the Riccati equation, suggesting that our method could be extended to non-BPS equations 
by inclusion of dynamics.
Such extension might shed more light onto the underlying boundary conformal theory,
an issue indicated in Ref.\cite{wieg}. 
The solutions we have obtained using this method are connected by the duality relation
$\l \rho(x)=\alpha-m(x)$, where $m(x)$ satisfies the BPS equation 
for  the  $\l'=1/\l$  case. Owing to the $su(1,1)$ dynamical symmetry these solutions  
correspond to the quasi-classical CFT duals of the giant gravitons on 
$AdS_5$ ($\l<1$) and on the sphere $S^5$ ($\l>1$) \cite{Cald,Bena,Cor,ber}.
Further, the method has enabled us to solve the equations which govern the dynamics of quantum excitations 
around the uniform and various $x-$dependent backgrounds.
These excitations described by the Hamiltonian in quadratic approximation
represent quantum corrections to the semiclassical solutions.
As an application of the results for quantum excitations 
on various $x-$dependent backgrounds,
one could determine the correlation functions,
the wave functionals of different states and the transition amplitudes among them.
On the other hand, one could evaluate the 
same quantities using random matrix theory \cite{br}
(at least for  $\l=1,1/2,2$) and then compare the results.

We have found the finite form of the $n$-soliton solution in the 
duality-based matrix
model, indicating the  complete integrability of this model. 
Owing to the origin of the duality-based matrix
model and to the properties of its semiclassical solutions, we
interpret this matrix model as a model of interacting giant gravitons having cubic interaction.
The BPS solutions for fields $\r(x)$ and $m(x)$ (\ref{spec2}) related by $\r(x) m(x)=\tilde{c}$ admit 
an interesting singular limit $\r(x) m(x)\to 0$;
at the places where one field is different from zero the other field is vanishing.
This "black/white" distribution  is a characteristic of 
the two-dimensional droplet model of electron gas \cite{iks}.
Finally, the singular limit of the  
$n$-soliton solution has motivated us to propose a 
realization of open-closed string duality. 
We have found the explicit mapping between the eigenstates of the collective-field Hamitonian
and the eigenstates of the Hamiltonian describing quantum mechanics of 
nonrelativistic fermions. 
The proposal could be made more precise
by expanding the relation (\ref{vv1}) between the wave functional of
the closed string excitations and  the wave function of dual  open
string excitations around one of the soliton solutions found.
We hope to address this issue in a future 
publication.

\section*{Acknowledgments}
This work was supported by the Ministry of Science, Education and Sports of the
Republic of Croatia.

\end{document}